\documentclass[prd,eqsecnum,twocolumn,amsfonts,amssymb]{revtex4}

\usepackage{graphicx}

\usepackage{bm}

\newcommand{\beq}{\begin{equation}}
\newcommand{\eeq}{\end{equation}}
\newcommand{\beqs}{\begin{eqnarray}}
\newcommand{\eeqs}{\end{eqnarray}}

\begin{document}

\title{On the Ultraviolet to Infrared Evolution of Chiral Gauge Theories}

\author{Thomas Appelquist$^a$ and Robert Shrock$^{a,b}$}

\affiliation{(a) \ Department of Physics, Sloane Laboratory \\
Yale University, New Haven, CT 06520}

\affiliation{(b) \ C. N. Yang Institute for Theoretical Physics and \\
Department of Physics and Astronomy, \\ 
Stony Brook University, Stony Brook, NY 11794\footnote{Permanent address}}

\begin{abstract}

We discuss the ultraviolet (UV) to infrared (IR) evolution of asymptotically
free chiral gauge theories. Various types of IR behavior are considered,
including confinement without spontaneous chiral symmetry breaking, and
formation of bilinear fermion condensates that preserve or dynamically break
the chiral gauge invariance.  We compare different schemes for the study of
this evolution and the resultant IR structure of the theories, including a
conjectured inequality based on the counting of degrees of freedom. We note
some new patterns of UV to IR evolution.

\end{abstract}


\maketitle

\section{Introduction}

Chiral gauge theories, in which the left- and right-handed chiral components of
the (massless) fermions couple differently to the gauge field, play a key role
in the standard model and in some efforts to extend the standard model. In the
latter case, chiral gauge theories without elementary scalar fields are often
explored. They can be constructed to be free of gauge anomalies, and with the
number of fermions limited, they are asymptotically free, typically becoming
strongly coupled in the infrared. They have been envisioned to play a role in
the generation of flavor hierarchies, dynamical supersymmetry breaking, and
quark and lepton substructure.

In this paper we revisit asymptotically free chiral gauge theories,
investigating their evolution from the ultraviolet to
the infrared.  Since the strength of the gauge interaction grows as the
renormalization-group (RG) scale $\mu$ decreases, there are,
{\it a priori}, several possible types of behavior that can set in.  Some early
studies of strongly coupled chiral gauge theories include
\cite{thooft79}-\cite{eppz}. We focus on theories that obey the 't Hooft
anomaly matching conditions \cite{thooft79}.

These allow, for example, confinement without breaking of the global or gauged
chiral symmetries. This would produce a set of massless, gauge-singlet, spin
1/2 composite fermions \cite{spinhalf}, although there is no dynamical basis
that we know of for the formation of these bound states. Another type of
infrared behavior, involving the formation of bilinear fermion condensates, is
also possible.  Because of the chiral nature of the gauge-fermion coupling,
these condensates typically break the gauge symmetry \cite{drs,etc}. Yet
another possibility is that the IR evolution may be controlled by a weak IR
fixed point, with neither confinement nor spontaneous chiral symmetry breaking.
The theory is then in the conformal window, occuring when the number of
massless fermions is sufficiently
large.

If the IR coupling is strong enough to trigger one of the first two types of
behavior, there remains little understanding of which might occur.  This
contrasts with vectorial gauge theories, where a number of general properties
have been established by a combination of continuum and lattice methods. For
chiral gauge theories, however, the presence of fermion doubling on the lattice
has made it challenging to construct a lattice implementation that could be
used to study their strong-coupling behavior. It seems worthwhile therefore to
re-examine the UV to IR evolution of these theories making use of various
continuum field-theoretic methods.

In addition to the 't Hooft anomaly matching conditions \cite{thooft79}, there
is the old most-attractive-channel (MAC) criterion for formation of bilinear
fermion condensates \cite{drs} based on single gauge-boson exchange. There also
exist schemes for constraining RG flow based on the counting of degrees of
freedom and the intuitive notion that flow to the IR should result in a
thinning of this count. Of the various implementations of this idea, the use of
the finite-temperature free energy to define the degree-of-freedom count
\cite{dfvgt,dfcgt} leads to a conjectured inequality that can usefully
constrain the IR behavior. For a class of chiral gauge theories, we examine the
consequences of this inequality and compare to the MAC criterion.

 In Section II, we review the general framework and methods
employed. In Section III, we describe the features of a simple chiral gauge
theory which we analyze with these methods. In Section IV, we describe a
general class of theories in which a set of massless fermions with vector-like
gauge couplings is added to the theory of Section III.  In Section V, we
summarize the possible phases of this general class. We present some
conclusions in Section VI.


\section{General Framework and Methods}


\subsection{Beta Function}

We focus on theories with gauge group $G={\rm SU}(N)$, and running
gauge coupling $\alpha(\mu) = g(\mu)^2/(4\pi)$.  The $\beta$ function,
$\beta_\alpha = d \alpha/dt$, where $dt = d\ln \mu$, has the loop expansion
\beq
\beta_\alpha = -2\alpha \sum_{\ell=1}^\infty b_\ell \, \Big (
\frac{\alpha}{4\pi} \Big )^\ell  \ ,
\label{beta}
\eeq
where we restrict to $b_1 > 0$ to insure asymptotic freedom. The coefficients
$b_\ell$ for $\ell \ge 3$ are scheme-dependent; in the $\overline{MS}$ scheme
they have been calculated up to four-loop order \cite{b3,b4}. If $b_2 < 0$, an
infrared zero can appear at two loops given by $\alpha_{IR,2\ell} = -4 \pi
b_{1} /b_{2}$ \cite{b2}. This a reliable result if $b_1/|b_2| << 1$, a
possibility depending on the number of massless fermions \cite{hl}. The
phenomena of main interest here, bilinear fermion condensate formation and/or
the appearance of massless composite fermions, are anticipated only at strong
coupling, where the loop expansion is not reliable.


\subsection{MAC Criterion}

If UV to IR evolution produces fermion condensates, an old scheme for
discriminating among possible breaking patterns is the
most-attractive-channel (MAC) criterion. For two chiral fermions in
representations $R_1$ and $R_2$ of the gauge group, a rough measure of the
likelihood of a condensation channel of the form $R_1 \times R_2 \to R_{cond.}$
is taken to be
\beq
\Delta C_2 = C_2(R_1) + C_2(R_2) -C_2(R_{cond.}) \ ,
\label{deltac2}
\eeq
where $C_2(R)$ is the quadratic Casimir invariant for the representation $R$
\cite{casimir}.

Since this is a measure of attractiveness arising from single-gauge-boson
exchange, its utility is uncertain, but the MAC criterion is that where
bilinear fermion condensates can occur and there are several possible channels,
the one that takes place has the largest value of $\Delta C_2$ \cite{drs}. (In
a vectorial gauge theory such as QCD, this implies that the only condensation
is $R \times \bar R \to 1$, preserving the gauge symmetry.)  A reasonable guess
then is that as $\mu$ decreases and $\alpha(\mu)$ increases, condensation will
first occur, in the MAC, when $\alpha(\mu)$ becomes of order $\alpha_{cr}$
given by
\beq
\frac{\alpha_{cr} \Delta C_2(R)}{2} \sim O(1) \ .
\label{alfcrit}
\eeq
%


\subsection{Conjectured Inequality Concerning UV versus IR Degrees of Freedom}

A general notion about RG flow to the IR, as first applied
to second-order phase transitions and critical phenomena, is the thinning of
degrees of freedom.  For two-dimensional conformal field theories (CFTs),
Zamolodchikov proved that the central charge, $c$, which may be considered to
count the degrees of freedom in the theory, decreases as a consequence of this
flow; that is, the new CFT to which the theory flows in the IR has a smaller
value of $c$ than the UV CFT \cite{zam}.

In four space-time dimensions, an approach using finite temperature $T$ as the RG scale
and the thermodynamic free energy $F(T)$ to count the degrees of freedom, can
provide significant constraints on the UV to IR flow of a quantum field
theory.  The quantity $f(T)$, given by
\beq
F(T) \equiv f(T) \, \frac{\pi^2}{90} \, T^4 \ ,
\label{f}
\eeq
is defined to count a single, massless bosonic degree of freedom as $1$ when
the theory is free. In this case,
\beq
f = 2 N_V + \frac{7}{4}N_F + N_S  \ ,
\label{fcount}
\eeq
where $N_V$ is the number of massless gauge fields, $N_F$ is the number of
massless chiral components of fermion fields, and $N_S$ is the number of scalar fields.  Thus, for an
asymptotically free, vectorial SU($N$) gauge theory with $N_f$ Dirac
fermions transforming according to the fundamental representation, $f_{UV} \equiv f(\infty) =2(N^2-1)+ (7/2) N_fN$.

A conjectured inequality \cite{dfvgt} is
that for an asymptotically free theory,
\beq
f_{IR} \equiv f(0) \leq f_{UV} \equiv
f(\infty) \, .
\label{inequality}
\eeq
No asymptotically free counterexample has yet been identified, although there
are theories for which $f(T)$ is not monotonic \cite{dfvgt}.  The inequality
can be directly implemented when there exists an IR effective field theory
(EFT) that is free or weakly interacting, allowing $f_{IR}$
to be computed perturbatively.  A weak IR
fixed point of the underlying gauge theory, for example, would make this
possible. In this case, $f_{IR} = f_{UV}$ + perturbative corrections, where the
leading ($O(\alpha_{IR})$) correction is negative.  So the inequality is
satisfied.

With a strong IR gauge coupling, a new, IR-free phase can emerge, consisting of
the massless, composite Nambu-Goldstone bosons (NGBs) associated with
condensate formation, or massless, gauge-singlet, composite fermions in the
case of chiral gauge theories. In these cases, the inequality can discriminate
simply among the various possibilities, its constraining power arising from the
fact that $f_{UV}$ is linear in fermion number, while $f_{IR}$, counting
massless composite degrees of freedom, is typically quadratic.  The inequality
was extended to chiral gauge theories in Ref. \cite{dfcgt}, and many examples
were discussed in Refs. \cite{dfvgt,dfcgt},\cite{ads}. A further possibility is
that the favored \ IR phase of a gauge theory is the one that not only
satisfies the inequality \ref{inequality}, but also minimizes $f_{IR}$.  


Another approach, following Ref. \cite{zam}, has investigated the possibility
that for \emph{some} quantity measuring the number of degrees of freedom in
four space-time dimensions, one could \emph{prove} either (i) that it is a
non-increasing function of the RG flow from the UV to the IR or (ii), a weaker
condition, that its value at a UV fixed point is greater than its value at an
IR fixed point. Some progress has been achieved by focusing on the coefficient
$a$, entering the trace of the energy-momentum tensor, $\langle \theta^\mu_\mu
\rangle = cW_{\kappa\lambda\mu\nu}W^{\kappa\lambda\mu\nu} - a E_4$, where
$W_{\kappa\lambda\mu\nu}$ is the Weyl tensor and $E_4$ is the four-dimensional
Euler density, whose integral is the Euler characteristic of the given
spacetime manifold \cite{atheorem,atilde}. For a free theory,
\beq
a=62N_V+\frac{11}{2}N_F+N_S \ ,
\label{a}
\eeq
where $N_V$, $N_F$, and $N_S$ denote the number of
massless vector, chiral fermion, and scalar fields. For the example of
an asymptotically free vectorial SU($N$) gauge theory
with $N_f$ massless Dirac fermions in the fundamental representation,
$a_{UV}=62(N^2-1)+11N_fN$. 

Both $a$ and $f$ are normalized to weight a single scalar field with
coefficient $1$. A peculiar feature of $a$ is the relative largeness of the
coefficient, $62$, emerging from a detailed computation, multiplying $N_V$ in
Eq. (\ref{a}).  By comparison, the coefficient $2$ multiplying $N_V$ in
Eq. (\ref{fcount}) which is applicable when the effective theory is weakly
coupled, simply counts the two transverse gauge boson degrees of freedom. In
the case of $a$, the large gauge boson term, which is present in the UV for an
asymptotically free theory, means that the inequality constraint $a_{IR} \leq
a_{UV}$ is not difficult to satisfy. 

Yet another approach to describing the IR properties of quantum field theories
involves compactification of the ${\mathbb R}^4$ space to ${\mathbb R}^3 \times
{\rm S}^1$, with periodic boundary conditions for all fields, and associated
analysis of topological properties \cite{poppitzunsal}. A degree-of-freedom
counting inequality could perhaps be developed in this framework. In the
present paper, we will focus on the the thermal inequality $f_{IR} \leq
f_{UV}$.  Although it is still conjectural, it makes a clear connection between
the Euclidean momentum scale and temperature in its weighting of coefficients
of $N_V$, $N_F$, and $N_S$, and, moreover, it can be quite constraining.


\subsection{Lattice Studies of Vectorial Theories}

Lattice studies have been carried out to investigate the IR behavior of many
vectorial, non-supersymmetric SU($N$) gauge theories. All results are so far
compatible with the inequality (\ref{inequality})
\cite{latticerefs}. Investigations of vectorial SU(2) gauge theories, currently
underway \cite{SU2}, are particularly interesting with respect to the
inequality. Here, the reality of the fermion representations leads to a larger
global symmetry, and typically therefore to a larger count of IR degrees of
freedom when this symmetry is broken.


\section{A Simple Theory}

We first review the UV to IR evolution of a familiar SU($N$) chiral gauge
theory \cite{by}. The chiral gauge symmetry forbids fermion mass terms in the
Lagrangian. Without loss of generality, we write all fermions as left-handed.

\subsection{Structure}

The fermion content consists of a symmetric, rank-2 tensor,
$\psi^{ab}_L \equiv \psi^{(ab)}_L$, together with $N+4$ copies (flavors) of the
conjugate fundamental representation, $\chi_{a,i,L}$, $i=1,...,N+4$. Here and
below, $a,b,..$ are SU($N$) gauge indices with $N \ge 2$, and $i,j,...$ are
copy indices. That is, the fermion content is:
\beq
 \psi^{(ab)}_L \ , \quad  \chi_{a,i,L} \ , \quad
i=1,...,N+4 \ .
\label{models}
\eeq

With a normalization convention in which the contribution of a left-handed
fermion in the fundamental representation of SU($N$) to the triangle anomaly is
1, the contribution of left-handed fermions in the rank-2 symmetric and
antisymmetric representation is $(N+4)$. Thus, the fermion content yields
theories that are free of chiral anomalies in gauged currents.

The one- and two-loop $\beta$ function coefficients are
\beq
b_1 = 3N - 2, ~~~~~ b_2 = \frac{1}{2}( 13N^2 - 30 N + 1 + 12N^{-1} )\, .
\label{b1,2sa}
\eeq
Because of
asymptotic freedom, the $f_{UV}$ function is given by
\beq
f_{UV} = 2(N^2-1) + \frac{7}{4}\bigg [ \frac{N(N + 1)}{2} + (N + 4)N \bigg ]
\, .
\label{fuv_sa}
\eeq

The classical global flavor symmetry group is ${\rm U}(1) \otimes {\rm
U}(N+4)_{\bar F}$.  Expressing ${\rm U}(N + 4)_{\bar F} = {\rm SU}(N + 4)_{\bar
F} \otimes {\rm U}(1)_{\bar F}$, we recall that the U(1) and U(1)$_{\bar F}$
are both rendered anomalous by instantons, but one can construct a linear
combination, denoted $\tilde {\rm U}(1)$, that is conserved in the presence of
instantons. Hence, the actual (non-anomalous) global flavor group of this
theory is
\beq
G_{f} =  {\rm SU}(N+4)_{\bar F} \otimes \tilde{\rm U}(1)
\label{gfs} \ .
\eeq


\subsection{Evolution}

\subsubsection{\bf Confinement without Chiral Symmetry Breaking}

A possibility is that the gauge interaction
confines and produces gauge-singlet, massless composite fermions. The
 't Hooft conditions, that the anomalies associated with the global
chiral flavor symmetries must be the same for the fundamental fields and for
the composites, are satisfied if the latter are the composite fermions
\beq
B_{[ij]} = \bar F_{a,i} S^{ab} \bar F_{b,j} - (i \leftrightarrow j)
\ ,
\label{bij_smodel}
\eeq
transforming according to the conjugate-antisymmetric, rank-2 tensor
representation of the global ${\rm SU}(N+4)$ symmetry group.

With only massless composite fermions in the spectrum, the low-energy effective
theory (EFT) consists of interaction operators only of dimension-6 and higher,
allowing $f_{IR}$ to be computed in the free theory:
\beq
f_{IR,SYM} = \frac{7}{4}\bigg [ \frac{(N+4)(N+3)}{2} \bigg ] \ .
\label{fir_s_sym}
\eeq
Hence, $f_{UV}-f_{IR,SYM}$ is non-negative for $N \ge 1.607$, that is, for all
physical values of $N$.


\subsubsection{\bf Chiral Symmetry Breaking}

An alternate possibility is that as $\mu$ decreases to some scale $\Lambda_N$,
the gauge interaction grows sufficiently strong to produce bilinear fermion
condensates.  The MAC is $S \times \bar F \to F$, with
\beq
\Delta C_2(S \times \bar F \to F) = C_2(S) = \frac{(N+2)(N-1)}{N} \ .
\label{deltac2_sfbarf}
\eeq
The condensate then has the form $\langle \psi^{ab \ T}_L C \chi_{b,i,L}
\rangle$.  Taking $a=1$ and $i=1$, it is
\beq
\langle \sum_{b=1}^N \psi^{1 b \ T}_L C \chi_{b,1,L} \rangle \, ,
\label{scondensateN}
\eeq
breaking the SU($N$) gauge symmetry to SU($N-1$) and the global ${\rm
SU}(N+4)_{\bar F}$ symmetry to ${\rm SU}(N+3)_{\bar F}$. The fermions in this
condensate, $\psi^{1b}_L$ and $\chi_{b,1,L}$, gain dynamical masses of order
$\Lambda_N$, and the $2N-1$ gauge bosons in the coset SU($N$)/SU($N-1$) gain
masses of order $g \Lambda_N$, where $g$ is the SU($N$) gauge coupling at the
scale $\Lambda_N$.

The residual SU($N-1$) gauge theory is then capable of self breaking to
SU($N-2$), and so on sequentially. Suppose, for simplicity, that $N=3$. After
the first breaking, there remains an SU($2$) gauge theory with three chiral
fermions in the symmetric tensor $\psi^{ab}_L$ ($a, b = 2,3$), along with
$18$ chiral fermions in $\chi_{b,i,L}$ ($b = 1,2,3$, $i = 2,...,7$).  The six
fermions with $b = 1$ are SU(2) gauge singlets, while the others comprise six
SU(2) doublets.  Note that the symmetric rank-2 tensor representation of SU(2)
is equivalent to the adjoint representation.

The SU(2) gauge theory can then fully self-break, perhaps at a lower scale. We
form a symmetric combination of two $\chi_{b,i,L}$, mixing SU(2) and flavor
labels as the chiral fermions that pair with the three $\psi^{ab}_L$, and, with
no loss of generality, we choose the flavor indices $i=2,3$ for these. The
condensate is thus
\beq
\langle \sum_{a,b=2}^3 \psi^{ab \ T}_L C \chi_{\{b,i \},L}\rangle \ ,
\label{lastscondensate}
\eeq
where $\chi_{\{b,i\},L}$ ($b=2,3, i = 2,3$) is the symmetrized two-index
combination. The remaining massless chiral fields are the $\chi_{b,i,L}$
($b=1$, $i=2,3$) and the single remaining, antisymmetrized $\chi_{[b,i],L}$
($b,i = 2,3$), forming an antisymmetric tensor under a global SU$^\prime$(3).
There are also the $12$ chiral fermions $\chi_{b,i,L}$ ($b = 1,2,3$, $i =
4,...,7$), transforming as an anti-fundamental under the global group
SU$^\prime$(3) $\times$ SU(4).

For general $N$, after all breakings, the
SU($N$) gauge group is fully broken, and the unbroken global symmetry
is \cite{ads}.
\beq
G_{f}' = {\rm SU}'(N) \otimes {\rm SU}(4) \otimes {\rm U}(1)' \, .
\label{gfsprime}
\eeq
There are $N(N-1)/2$ massless elementary chiral fermions transforming as the
antisymmetric tensor representation of ${\rm SU}'(N)$ and $4N$ massless
elementary chiral fermions transforming as the anti-fundamental representation
of ${\rm SU}'(N) \otimes {\rm SU}(4)$, for a total of $N(N+7)/2$ massless
chiral fermions..  There are also $8N+1$ composite NGBs. This is the difference
between the number of generators for the initial local-plus-global symmetry
group, ${\rm SU}(N) \otimes G_{f}$, and the final (global) symmetry group
$G_{f}'$, minus the $N^2-1$ absorbed to give mass to the gauge bosons of the
original SU($N$) gauge theory.

The EFT describing these massless degrees of freedom includes interaction operators only of
dimension $d > 4$, allowing $f_{IR}$ to be computed in the free theory. One
finds
\beq
f_{IR,S \times \bar F} = 8N+1 + \frac{7}{4}\bigg [ \frac{N(N-1)}{2} + 4N \bigg ] \ .
\label{fir_s_sfbar}
\eeq
Hence, $f_{UV} - f_{IR, S \times \bar F}$ is nonnegative for $N \geq
2.0558...$.  We also note that $f_{IR,SYM} < f_{IR,S \times \bar F}$ for all
integer $N$, possibly favoring massless composite fermion formation.

Before discussing a generalization of this theory, we note that one could also
consider a theory with an SU($N$) gauge group with $N \ge 5$ containing a
massless fermion content consisting of a rank-2 antisymmetric reprepresentation
$\psi^{[ij]}_L$ and $N-4$ copies of the conjugate fundamental representation,
denoted $\bar F$ \cite{dfcgt,ads}.  This theory is free of anomalies in gauged
currents and satisfies the 't Hooft matching conditions. As with the present
theory, there are several possible types of IR behavior, including (i)
confinement without spontaneous chiral symmetry breaking, yielding massless
composite fermion and (ii) formation of fermion condensates breaking gauge and
global symmetries.  In the case (ii), for SU(5), the MAC is $A \times A \to
\bar F$; for SU(6) there are two equally attractive MACs, $A \times A \to \bar
A$ and $A \times \bar F \to F$, while for SU($N$) with $N \ge 7$, the MAC is $A
\times \bar F \to F$.


\section{A General Class of Theories}

We now consider a one-parameter family of chiral gauge theories formed by the
addition to the theory of Eq. (\ref{models}) of a vectorlike sector consisting
of $p$ copies of the pair of fermions $\{F + \bar F \}$ \cite{by}.


\subsection{Structure of the General Class}

The fermionic content is
\beqs
 \quad & & \psi^{(ab)}_L \ , \quad \chi_{a,i,L},
\ i=1,...,N+4+p \, \cr\cr
& & \omega^a_{j,L} \ , j=1,...,p  \, .
\label{spmodel}
\eeqs
The classical global flavor symmetry group is ${\rm U}(1) \otimes {\rm
U}(N+4+p)_{\bar F} \otimes {\rm U}(p)_{F}$.  It can equivalently be written as
${\rm U}(1) \otimes [{\rm SU}(N+4+p)_{\bar F} \otimes {\rm U}(1)_{\bar F}]
\otimes [{\rm SU}(p)_{F} \otimes {\rm U}(1)_{F}]$.  The U(1) factor groups are
rendered anomalous by SU($N$) instantons, but one can form two linear
combinations that are invariant.  Hence, the actual (non-anomalous) global
flavor group is
\beq
G_{f}^{p}={\rm SU}(N+4+p)_{\bar F} \otimes {\rm SU}(p)_{F}
\otimes \tilde{\rm U}(1) \otimes \tilde{\rm U}(1)' \, .
\label{gfsp}
\eeq

The one-loop coefficient in the $\beta$ function
is
\beq
b_1^{p} = 3N - 2 - \frac{2p}{3} \, .
\label{b1_spapmodel}
\eeq
In the asymptotically free range, $p < \frac{9N}{2} - 3$,
one can identify the free UV degrees of freedom, with the result
\beqs
& & f_{UV}^{p} = f_{UV} + \frac{7}{4}(2pN) \cr\cr
& = & 2(N^2-1) + \frac{7}{4} \bigg [ \frac{N(N + 1)}{2}
+ (N + 4 + 2p)N \bigg ] \ . \cr\cr
& &
\label{fuvsa}
\eeqs
The two-loop $\beta$-function coefficient is
\beq
b_2^{p} = \frac{13N^2}{2} - 15N + \frac{1}{2} + 6N^{-1} +
p \Big ( -\frac{13 N}{3} + N^{-1} \Big )\, .
\label{b2_spapmodels}
\eeq
%


\subsection{Evolution of the General Class}

\subsubsection{\bf Non-Abelian Coulomb Phase}

For a range of $p$ values below $9/2 - 3$, the two-loop $\beta$
function has a non-trivial zero (an IR fixed point) at
\beq
\alpha_{IR,2\ell}^{p} = \frac{8\pi N (9N - 6 -2p)}
{-39N^3 + 90N^2 -3N - 36+p(26N^2-6)} \, ,
\label{alfir_2loop_spap}
\eeq
a reliable result providing $9/2 - 3/N - p/N \ll 1$ so that the fixed point is
weak. In this range,
\bigskip
\beq
f_{IR,NAC}^{p} = f_{UV}^{p} + {\rm perturbative ~ corrections}\, ,
\label{fir conf}
\bigskip
\eeq
with the leading correction small and negative. Thus $f_{UV}^p >
f_{IR,NAC}^{p}$.

The fixed-point strength increases monotonically with decreasing $p$, becoming
inaccessible via perturbation theory when $9/2 - 3/N - p/N = O(1)$. It is
expected to reach a strength necessary for confinement and/or symmetry breaking
at some critical value $p_{cr} = O(N)$, not precisely known.

In the following we assume that there exists a $p_{cr}$ below which the theory
leaves the non-Abelian Coulomb phase. There are then several possibilities for
its IR behavior.


\subsubsection{\bf Confinement With No Chiral Symmetry Breaking}

 One possibility is confinement without spontaneous chiral symmetry
breaking producing a set of massless composite fermions
transforming according to (i) the conjugate of the
antisymmetric rank-2 tensor representation of the SU($N+4+p$) global symmetry
group and the singlet representation of the SU($p$) group; (ii) the fundamental
representation of SU($N+4+p$) and SU($p$); and (iii) the singlet representation
of SU($N+4+p$) and the conjugate of the symmetric representation of SU($p$)
\cite{by}.

The EFT consists of interaction operators only of dimension $6$ and higher, so that
the free theory may be used to determine $f_{IR,SYM}^{p}$. Thus,
\beqs
f_{IR,SYM}^{p} & = & \frac{7}{4} \bigg [ \frac{(N+4+p)(N+3+p)}{2} + p(N+4+p)
\cr\cr & + &   \frac{p(p+1)}{2} \bigg ]
\label{fir_sp_sym} \, .
\eeqs
Hence, $f_{UV}^{p} - f_{IR,SYM}^{p}$ is nonnegative providing
\beq
p \leq -2 +  \bigg [ \frac{15N^2+7N+6}{14} \bigg ]^{1/2}\, .
\label{pzero_df_sp_sym}
\eeq
The inequality (\ref{inequality}) thus predicts that for $p$ in
this range, the interaction can confine and
produce massless composite fermions, whereas it is forbidden for larger
$p$-values.


\subsubsection{\bf Chiral Symmetry Breaking in the $S \times \bar F \to F$
Channel}

\bigskip

There are various possibilities for IR behavior that involve
dynamical chiral symmetry breaking.  These can be classified according to
several criteria, including

\bigskip

(i) their attractiveness measure $\Delta C_2$,

\bigskip

(ii) whether they also break the SU($N$) gauge symmetry,

\bigskip

(iii) the value of $f_{IR}$, which depends on the details of the sequential
formation of condensates and breaking of global (and possibly gauge) symmetries
at corresponding scales.

\bigskip

The most attractive channel (MAC) for fermion condensation is the
$S \times \bar F \to F$ channel, with measure $\Delta C_2(S
\times \bar F \to F)$. The attractiveness for this channel may be compared with
that for the channel $F \times \bar F \to 1$.  The difference between the
$\Delta C_2$ values for these channels is
\beqs
\Delta C_2(S \times \bar F \to F) & -& \Delta C_2(F \times \ \bar F \to 1) = C_2(S) - 2C_2(F) \cr\cr
& = & \frac{N-1}{N} > 0 \ . \cr\cr
&&
\label{difdif_sp}
\eeqs

Thus, if the gauge interaction causes the formation of fermion condensates for
$p < p_{cr}$, rather than confining without chiral symmetry breaking, the MAC
criterion points to condensation in the $S \times \bar F \to F$ channel. In
this case, we denote the scale where this condensation occurs as
$\Lambda_N$. Here, the gauge symmetry is broken from SU($N$) to SU($N-1$).

If $N \ge 3$, the same MAC argument implies that in the effective theory for
scales below $\Lambda_N$, there will again be condensation in the $S \times
\bar F \to F$ channel at a lower scale $\Lambda_{N-1}$, breaking SU($N-1$) to
SU($N-2$), and so forth.  At each scale, the fermions involved in the
condensate will gain dynamical masses. Since the $N+4$ $\chi_{a,i,L}$,
$i=1,...,N+4$, suffice to break the SU($N$) symmetry completely, the additional
$p$ vectorlike pairs of fermions $F, \ \bar F$ remain in the IR.  Hence, the
number of massless chiral components of fermions in the IR EFT is calculated by
simply adding these to the number $N(N-1)/2+4N$ in the $p=0$ theory, for a
total of $N(N-1)/2 + 4N + 2pN = N(N+7+4p)/2$.  Correspondingly, the final
global symmetry group is
\beq
G^{'p}_{f} = {\rm SU}'(N) \otimes {\rm SU}(4+p) \otimes {\rm SU}(p) \otimes
{\rm U}(1)'_{p} \otimes {\rm U}(1)''_{p} \ .
\label{GMAC}
\eeq
The number of NGBs is $2N(4+p)+1$.  This is the difference between the number
of generators of the initial local+global symmetry group ${\rm SU}(N) \otimes
G_{f}^{p}$, and the final (global) symmetry group $G^{'p}_{f}$, minus the
$N^2-1$ absorbed to give masses to the gauge bosons of the original SU($N$)
gauge theory.

The IR EFT again consists of interaction operators only of dimension $d > 4$, so that
the free theory determines $f_{IR}$:
\beq
f_{IR,S \times \bar F}^p=2N(4+p)+1+\frac{7}{4}\Big [ \frac{N(N-1)}{2}+4N+2pN\Big ]\, ,
\eeq
behaving linearly with p as does $f_{UV}^{p}$, but with a larger slope. The
linearity of the NGB count with respect to $p$ is notable, requiring a large
unbroken global symmetry.

For this UV to IR evolution, $f_{UV}^p - f_{IR,S \times \bar F}^p$ is always
negative for $N=2$, but for $N\geq 3$, it is nonnegative (in accord with the
conjectured inequality) for
\beq
p \le \frac{15N^2-25N-12}{8N}
\label{pmax_sp_sfbar}
\eeq
For $N=3$, for example, $p \leq 2$.


\subsubsection{\bf Chiral Symmetry Breaking in the $F \times \bar F \to 1$
Channel Followed by Confinement with no Further Symmetry Breaking}

Given the uncertainties in the strong-coupling physics involved, it is
important to consider other possible condensation channels.  A natural one is
the $F \times \bar F \to 1$ channel, involving the $p$ fermions in the
vectorlike subsector of the theory.  Although it is less attractive than the $S
\times \bar F \to F$ channel, it has the important feature that it does not
break the SU($N$) gauge symmetry.  The condensate for this channel is
\beq
\langle \chi^T_{a,i,L} C \omega^a_{i,L} \rangle \ , \quad i = N+4+1,...,N+4+p
\label{ffbarcondensate}
\eeq
This condensation pattern was studied in \cite{ads}.  We denote the scale where
it occurs as $\Lambda_V$. It breaks the $G_{f}^{p}$ global symmetry to
\cite{ads}
\beq
G_{f}^{\prime p} =  {\rm SU}(N+4)_{\bar F} \otimes {\rm SU}(p)_V \otimes {\rm U}(1)_1' \otimes
{\rm U}(1)_2' \ ,
\label{gfsp1}
\eeq
leading to $p(2N+p+8)$ NGBs. The fermions involved in the condensate
(\ref{ffbarcondensate}) get dynamical masses of order $\Lambda_V$, and the
low-energy effective field theory is then the $p=0$ theory.

The further infrared evolution of this theory has been reviewed in Section
III. The possibility of confinement without further chiral symmetry breaking is
favored by $f_{IR}$ minimization over further chiral symmetry breaking (along
with gauge symmetry breaking).  Massless composites are then formed, the EFT
includes interaction operators only of dimension $d > 4$, and
the function $f_{IR}$ is \cite{ads}
\beq
f_{IR,F \times \bar F, SYM}^{p} = p(2N+p+8) + \frac{7}{4}
\Big [ \frac{(N+4)(N+3)}{2} \Big ]
\label{firspffbarconfine}
\eeq

Hence, $f_{UV}^p - f_{F \times \bar F, SYM}^{p}$ is nonnegative, and in accord
with the conjectured inequality (\ref{inequality}) providing
\beq
p \le \frac{1}{4} \Big [ 3N-4 + \sqrt{69N^2-68N+56} \ \Big ]
\label{pbound_sp_ffbarsym}
\eeq
For $N=3$, for example, $p \le 3.687$, allowing
the physical values $p=0, \ 1, \ 2, \ 3$.


\subsubsection{\bf Chiral Symmetry Breaking in the $F \times \bar F \to 1$
Channel Followed by $S \times \bar F \to F$ Chiral Symmetry Breaking}

Finally, if after the condensation of the $p$ vector-like pairs in the channel
$F \times \bar F \to 1$, the residual, $p=0$ theory breaks the chiral and gauge
symmetries via $S \times \bar F \to F$, as described in Section IIIB2, then the
gauge symmetry is broken and the remaining global symmetry is
\beq
G_{f}^{\prime p} = {\rm SU}^{\prime}(N) \otimes {\rm SU}(4)\otimes
 {\rm SU}(p)_V \otimes {\rm U}(1)_1' \otimes {\rm U}(1)_2' \, .
\label{gfsp2}
\eeq
The IR EFT again consists of interaction operators only of dimension $d > 6$. The
resultant $f_{IR}$ counts the massless degrees of freedom, and is
\beqs
f^{p}_{IR,F \times \bar F, S \times \bar F} & = & (2pN +p^2 + 8p) + (8N+1)
\cr\cr
& + & \frac{7}{4}[\frac{1}{2}N(N-1) +4N] \, .
\eeqs
Hence, $f_{UV}^p - f^{p}_{IR,F \times \bar F, S \times \bar F}$ is nonnegative
for
\beq
p \le \frac{1}{4}\bigg [ 3N-16 + \sqrt{69N^2-196N+208} \ \bigg ]
\label{pbound_sp_ffbar_to_sfbar}
\eeq
For $N=3$, for example, $p \le (-7+\sqrt{241} \ )/4 = 2.131$,
allowing the physical values $p=1, \ 2$.


\section{Summary of the General Class of Theories}

The class of theories considered here has several possibilities for UV to IR
evolution. For $p/N$ sufficiently close to the upper limit $9/2 - 3/N$:

\bigskip

(i) to a non-Abelian Coulomb phase with an IR fixed point of the gauge
coupling and particle content consisting of the elementary gauge fields and
fermions;

\bigskip

\noindent Then, for $p \leq$ some critical value $ p_{cr}(N)$, there are
several strong-coupling possibilities:

\bigskip

(ii) to a phase with confinement, massless composite fermion formation and no
gauge or chiral symmetry breaking;

\bigskip

(iii) to a phase with sequential condensation in the respective $S \times \bar
F \to F$ (MAC) channels, breaking the SU($N$) gauge symmetry completely,
leaving a set of massless elementary fermions and massless composite NGBs
in the infrared EFT;

\bigskip

(iv) to a phase with condensation of the $p$ vectorlike fermions in the channel
$F \times \bar F \to 1$ followed by confinement with massless composite fermion
formation, no further chiral symmetry breaking, and no gauge-symmetry breaking,
so that the IR EFT consists of the massless
composite fermions together with massless NGBs;

\bigskip

(v) to a phase with condensation of the $p$ vectorlike fermions in the channel
$F \times \bar F \to 1$, followed by condensation in the $S \times \bar F \to
F$ channel, again breaking the SU($N$) gauge symmetry completely, so that the
IR particle content consists of massless NGBs and massless elementary fermions

\bigskip

\noindent We note that each of these phases has either
elementary or composite massless fermions, but not both. 

\subsection{${ \bf N = 3}$}

To discuss these possibilities further, we first consider the case $N=3$. In Fig. 1,
we plot $f_{IR}$ for each of the strong coupling phases, along with
$f_{UV}$. For $p/N$ near $3.5$, the theory is in the non-Abelian Coulomb phase
(i), with $f_{IR,NAC} \leq f_{UV}$, the difference being computable in
perturbation theory. There is no confinement and no symmetry breaking. This
$f_{IR}$ curve is not shown explicitly in Fig. 1.

As $p$ is decreased, other IR phases become possible. With $p$ still relatively
large, the MAC phase (iii) (red) with a large unbroken global symmetry
(\ref{GMAC}) and no composite fermions has the smallest $f_{IR}$, but here
$f_{IR,S\times \bar F} > f_{UV}$. For smaller $p$, when $f_{IR,S\times \bar F}
\leq f_{UV}$, this phase does not give the smallest $f_{IR}$.

As $p$ is decreased further, the first phase to allow $f_{IR} \leq f_{UV}$, for
$p \leq 3.69$, is phase (iv) (dark blue) with condensation of the $p$
vector-like fermions followed by confinement with massless composite fermion
formation. There is no gauge symmetry breaking. Throughout the range $p \leq
3.69$, $f_{IR}$ for this phase is the smallest among the strong coupling
phases, but the breaking channel is not the MAC.

As $p$ is decreased still further, other phases are allowed by the inequality
$f_{IR} \leq f_{UV}$.  In particular, phase (iii) (red) involving sequential
breaking in the MAC channel $S \times \bar F \to F$, breaking the gauge
symmetry completely, satisfies the inequality for $ p \leq 2$, although it does
not give the smallest $f_{IR}$.

Phase (v) (black) is compatible with the inequality for $p \leq 2.131$, but it
never minimizes $f_{IR}$.  Phase (ii) (green), with confinement but no symmetry
breaking, respects the inequality for $p \leq 1.4$, but leads to the smallest
$f_{IR}$ only when $p=0$.

A plot of $f_{IR}$ values for $N > 3$ is similar.


\subsection{{\bf Large-N Limit} }

It interesting to consider the limit $N \rightarrow \infty$ \cite{eppz} with $
r \equiv p/N$ and $\xi(\mu) = \alpha^{2}(\mu)N$ held fixed. The upper bound on
$r$ from asymptotic freedom is $r < 9/2$, and the two-loop IR fixed point is at
\beq \xi_{IR,2\ell} = \frac{8\pi(9-2r)}{13(2r-3)}\, .  \eeq The reduced
quantity $ \bar{f} \equiv f/N^2$ is finite in this limit. We have
\beq
\bar f_{UV} = \frac{37}{8}+\frac{7r}{2} \ .
\label{fbaruvsap}
\eeq
The theory is in the non-Abelian Coulomb phase (i) for $r$ near $9/2$, with
$\bar f_{IR,NAC}$ perturbatively below $\bar f_{UV}$.

Phase (ii) with confinement, massless-composite-fermion formation and no
symmetry breaking, gives
\beq
\bar f_{IR,sym} = \frac{7}{8}(1+4r+4r^2) \, ,
\label{fbarir_sp_sym_lnn}
\eeq
satisfying the inequality for
$ r \leq 15/14$.

Phase (iii), with sequential condensation in the respective $S \times \bar
F \to F$ (MAC) channels, gives
\beq
\bar f_{IR,S \times \bar F} = \frac{7}{8}+\frac{11r}{2} \, ,
\label{fbar_sp_sfbar_lnn}
\eeq
satisfying the inequality for
$r \leq 15/8$.

Finally, with condensation of the $p$ vectorlike fermions in the channel $F
\times \bar F \to 1$, followed by either confinement without further chiral
symmetry breaking (phase (iv)) or further sequential condensation in the $S
\times \bar F$ channels (phase(v)), we have
\beq
\bar f_{IR,F \times \bar F, SYM} =
\bar f_{IR,F \times \bar, S \times \bar F} =
 r(2+r)+\frac{7}{8} \, .
\label{fbar_ir_sp_ffbarconfine_lnn}
\eeq
Interestingly, $f_{IR}$ is the same for these two phases even though the term
$7/8$ counts composite fermions in one case and residual elementary fermions in
the other.  The inequality is stisfied for $r < (3+\sqrt{69})/4 $.

In Fig. \ref{fplotlnn} we plot $\bar f$ for the various phases as functions of
$r$.  While the two curves for phases (iv) and (v) have collapsed to one, the
qualitative picture is otherwise similar to the finite-$N$ case.

The common curve for phases (iv) and (v) satisfies the inequality for the
largest $r$ range, and provides the smallest $f_{IR}$ throughout this
range. However, as in the finite-$N$ case, the condensation channel for these
phases is not the MAC, this being phase (iii) with sequential condensation in
the channel $S \times \bar F \to F$.

\bigskip
\section{Discussion and Conclusions}

We have examined a class of ${\rm SU}(N)$ chiral gauge theories including $p$
additional, vector-like fermions. In addition to a weak-coupling, non-Abelian
Coulomb phase (i), present for large enough $p/N$, we have described a set of
four possible strong-coupling phases. We have classified these phases with
reference to the degree-of-freedom-counting inequality $f_{IR} \leq f_{UV}$.

One phase, (ii), breaks no symmetries and includes massless composite
fermions. Another, phase (iii), new to this paper, involves sequential breaking
in the maximally-attractive (MAC) channel, breaks the gauge symmetry entirely,
and leaves a large unbroken global symmetry. This phase, together with two
others, (iv), and (v), involves chiral symmetry breaking through bilinear
fermion condensation, leaving unabsorbed massless Nambu-Goldstone bosons. Phase
(iii), in which the SU(N) gauge symmetry is broken sequentially, is the MAC
among these three.  Two phases, (ii) and (iv), involve the formation of
massless composite fermions, and leave the gauge symmetry unbroken. Two others,
(iii) and (v), break it completely, and leave residual massless, elementary
fermions.

Phase (iv), in which the vector-like fermions condense among themselves and the
chirally coupled fermions form massless composite fermions, leaving the SU(N)
gauge symmetry unbroken, satisfies the inequality $f_{IR} \leq f_{UV}$ for the
largest range of $p/N$ values, and minimizes $f_{IR}$ throughout this
range. But it is not the MAC for condensation. (In the limit $N \rightarrow
\infty$ with $ r \equiv p/N$ held fixed, phases (iv) and (v) give the same
value for $f_{IR}$.)  Phase (ii), with no symmetry breaking and
massless-composite-fermion formation, satisfies the inequality for small $p/N$
values, but minimizes $f_{IR}$ only at $p=0$. These results are all shown in
Figs. 1 and 2.

This survey of four possible strong-coupling phases is based on the conjectured
inequality $f_{IR} \leq f_{UV}$, along with the (possibly unreliable) MAC
criterion, deriving from single-gauge-boson exchange. Phase (iii), involving
MAC condensation, satisfies the inequality only for small $p/N$, and when it
does, it does not minimize $f_{IR}$.  In the absence of strong-coupling
computations or proven, restrictive constraints in addition to 'tHooft anomaly
matching, it remains unknown which of the strong-coupling phases is realized
for various values of $p$. While there is no dynamical basis that we know of
for the formation of massless composite fermions in chiral gauge theories, as
in phases (ii) and (iv), the formation of massless Nambu-Goldstone bosons, as
in phases (iii) and (v), can be realized in various dynamical schemes. It is
also an approximate feature of real-world QCD.  The quantity $f_{IR}$ for
phases (iii) and (v) is shown in the red and black curves of Figs. 1 and 2.

This discussion of possibilities can perhaps provide a helpful template for the
further study of the class of chiral gauge theories considered here, as well as
other theories.


This research was partially supported by the grants DE-FG02-92ER-40704 (T.A.)
and NSF-PHY-09-69739 (R.S.). One of us (T.A.) would like to acknowledge the
hospitality of the Aspen Center for Physics while this paper was being
completed.


\begin{figure}
  \begin{center}
    \includegraphics[height=16cm,width=14cm]{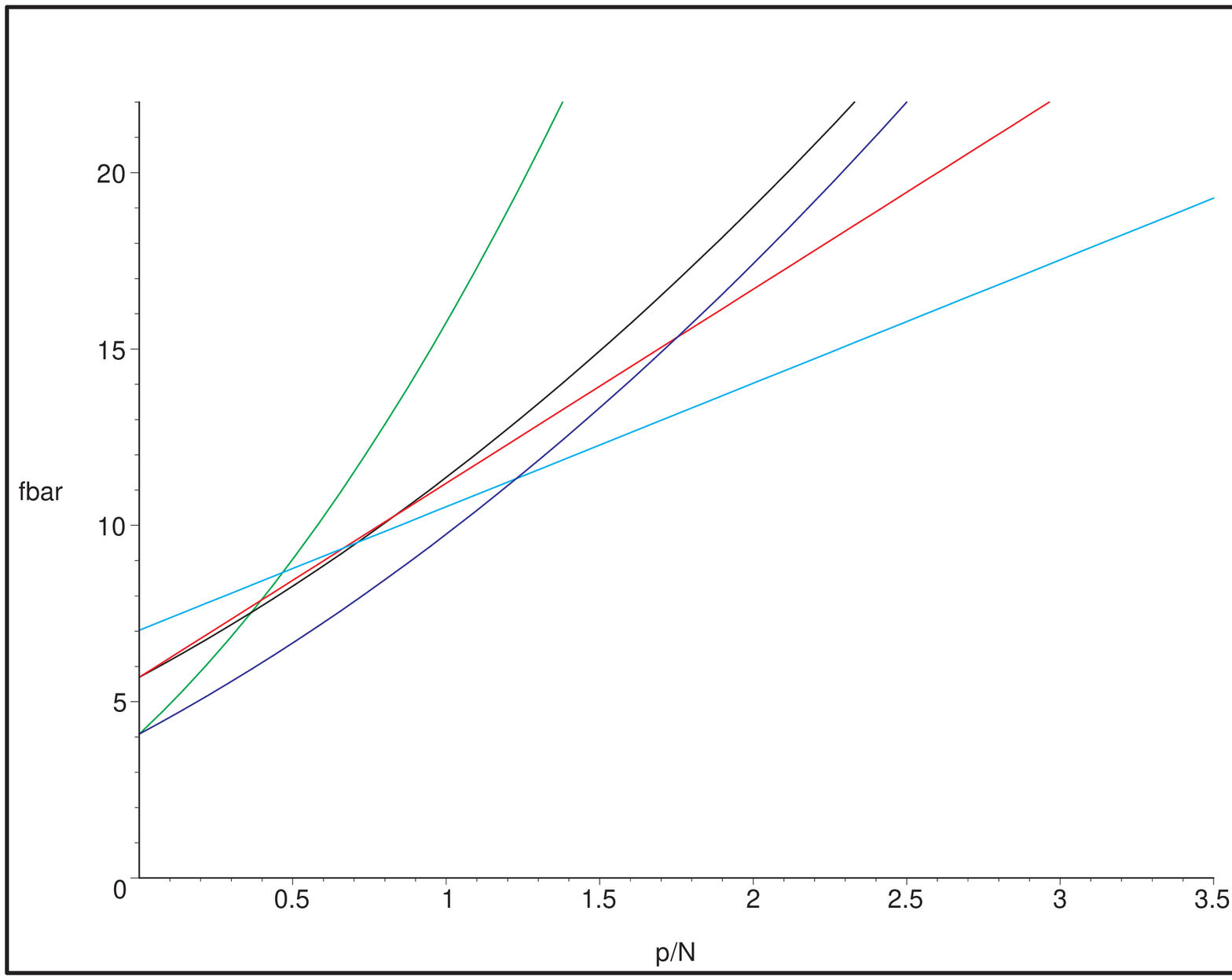}
  \end{center}
\caption{\footnotesize{Plot of $\bar f \equiv f/N^2$ for various types of IR
behavior, as functions of $p/N$, for the case $N=3$.  The curves (with color in
the online version) are ${\bar f}_{UV}^p$ (light blue), (ii) ${\bar
f}_{IR,SYM}^p$ (green), (iii) ${\bar f}_{IR,S \times \bar F}^p$ (red), (iv)
${\bar f}_{IR,F \times \bar F, SYM}^{p}$ (dark blue), and (v) ${\bar
f}^{p}_{IR,F \times \bar F, S \times \bar F}$ (black).  The curve for
$f_{IR,NAC}$ in phase (i), not shown, is perturbatively close to the curve for
${\bar f}_{UV}^p$. At $p/N=1.2$, the curves, from bottom to top, are (iv), (UV),
(iii), (v), and (ii).  }}
\label{fplot}
\end{figure}
%

%
\begin{figure}
  \begin{center}
    \includegraphics[height=16cm,width=14cm]{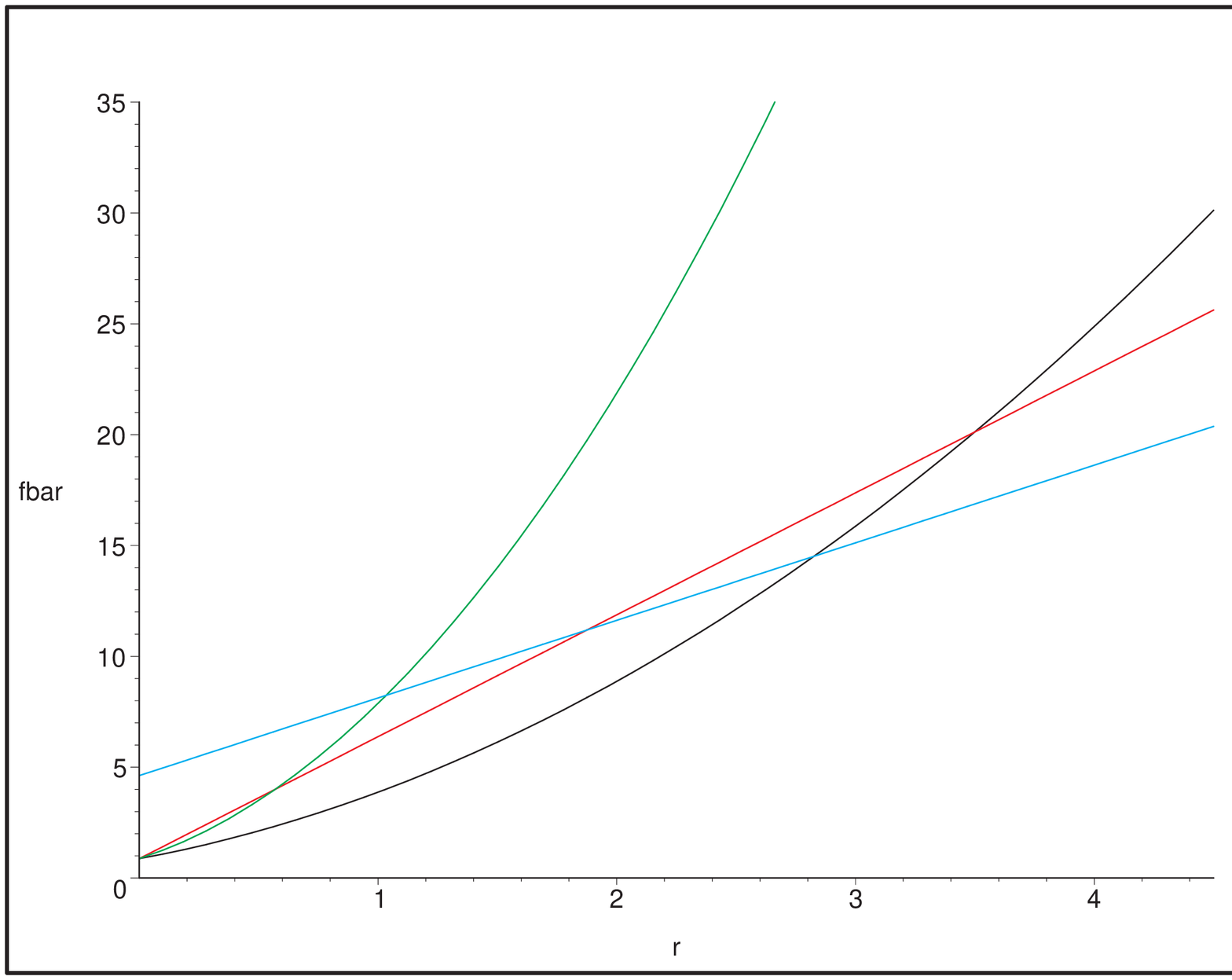}
  \end{center}
\caption{\footnotesize{Plot of $\bar f \equiv f/N^2$ for various types of IR
behavior, as functions of $r \equiv p/N$, in the large-$N$ limit (with color in
the online version). The curves are ${\bar f}_{UV}^p$ (light blue),
(ii) ${\bar f}_{IR,SYM}^p$ (green), (iii) ${\bar f}_{IR,S \times \bar F}^p$
(red), (iv and v) ${\bar f}_{IR,F \times \bar F, SYM}^{p} = {\bar f}^{p}_{IR,F
\times \bar F, S \times \bar F}$ (black).  The curve for $f_{IR,NAC}$ in phase
(i), not shown, is perturbatively close to the curve for ${\bar f}_{UV}^p$. At
$r=1.2$, from bottom to top, the curves are (iv and v), (iii), (UV), and
(ii).}}
\label{fplotlnn}
\end{figure}


\begin{thebibliography}{99}

\bibitem{thooft79}
G. 't Hooft, in {\it Recent Developments in Gauge Theories, 1979 Carg\`ese
Summer Institute} (Plenum, New York, 1980), p. 135.

\bibitem{spinhalf}
%
S. Weinberg and E. Witten, Phys. Lett. B {\bf 96}, 59 (1980).

\bibitem{drs}
S. Dimopoulos, S. Raby, and L. Susskind, Nucl. Phys. B {\bf 169}, 373 (1980);
S. Raby, S. Dimopoulos, and L. Susskind, Nucl. Phys. B {\bf 173}, 208 (1980).

\bibitem{g78}
H. Georgi, Nucl. Phys. B {\bf 266}, 274 (1986).

\bibitem{by}
I. Bars and S. Yankielowicz, Phys. Lett. B {\bf 101}, 159 (1981).

\bibitem{eppz}
E. Eichten, R. D. Peccei, J. Preskill, and D. Zeppenfeld, Nucl. Phys. B
{\bf 268}, 161 (1986).

\bibitem{etc}
%
See, e.g., T. Appelquist and
R. Shrock, Phys. Lett. B {\bf 548}, 204 (2002); Phys. Rev. Lett. {\bf 90},
201801 (2003); T. Appelquist, M. Piai, and R. Shrock, Phys. Rev. D {\bf 69},
015002 (2004); 
T. A. Ryttov and R. Shrock, Phys. Rev. D {\bf 81}, 115013 (2010).

\bibitem{dfvgt}
T. Appelquist, A. Cohen, and M. Schmaltz, Phys. Rev. B {\bf 60}, 045003
(1999).

\bibitem{dfcgt}
T. Appelquist, A. Cohen, M. Schmaltz, and R. Shrock, Phys. Lett. B {\bf 459},
235 (1999).

\bibitem{b3}
O. V. Tarasov, A. A. Vladimirov, and A. Yu. Zharkov, Phys. Lett. B {\bf 93},
429 (1980); S. A. Larin and J. A. M. Vermaseren, Phys. Lett. B {\bf 303}, 334
(1993).

\bibitem{b4}
T. van Ritbergen, J. A. M. Vermaseren, and S. A. Larin, Phys. Lett. B {\bf
 400}, 379 (1997).

\bibitem{b2}
W. E. Caswell, Phys. Rev. Lett. {\bf 33}, 244 (1974);
D. R. T. Jones, Nucl. Phys. B {\bf 75}, 531 (1974).

\bibitem{hl}
Higher-loop corrections have been studied in
E. Gardi and M. Karliner, Nucl. Phys. B {\bf 529}, 383 (1998);
E. Gardi and G. Grunberg, JHEP 03, 024 (1999);
T. A. Ryttov and R. Shrock, Phys. Rev. D {\bf 83}, 056011 (2011);
C. Pica and F. Sannino, Phys. Rev. D {\bf 83}, 035013 (2011).

\bibitem{casimir}
%
We use standard notation and normalization for Casimir and trace invariants.
In particular, $C_A \equiv C_2(G)$, and $C_f \equiv C_2(R)$
and $T_f \equiv T(R)$ for a fermion representation $R$.

\bibitem{zam}
A. B. Zamolodchikov, Pis'ma Zh. Eksp. Fiz. {\bf 46}, 129 (1987)
[Sov. Phys. JETP Letts. {\bf 46}, 160 (1987)].

\bibitem{ads}
T. Appelquist, Z. Duan, and F. Sannino, Phys. Rev. D {\bf 61}, 125009 (2000).

\bibitem{atheorem}
Z. Komargodski and A. Schwimmer, JHEP 1112, 099 (2011), arXiv:1107.3987;
Z. Komargodski, JHEP 1207, 069 (2012), arXiv:1112.4538.

\bibitem{atilde}
%
Actually, the quantity which has been argued to satisfy the local
monotonicity relation is a quantity $\tilde a$ that differs from $a$ along an
RG trajectory, but coincides with $a$ at RG fixed points \cite{atheorem}, as
reviewed, e.g., in Y. Nakayama, arXiv:1302.0884. However, we do not pursue the
implications of $\tilde a$ monotonicity here.

\bibitem{poppitzunsal}
E. Poppitz and M. \"Unsal, JHEP 0909, 050 (2009). 

\bibitem{latticerefs}
%
See talks at the workshop Lattice Meets Experiment 2012: Beyond the Standard
Model, Univ. of Colorado, Oct. 2012, at
 http://www-hep.colorado.edu/tilde schaich/lat-exp-2012; and Lattice 2013,
July, 2013, at www.lattice2013.uni-mainz.de.

\bibitem{SU2}
F. Bursa, L. Del Debbio, L. Keegan, C. Pica and T. Pickup, Phys. Lett. B
696, 374 (2011) [arXiv:1007.3067];
[35] T. Karavirta, J. Rantaharju, K. Rummukainen and K. Tuominen, JHEP
1205, 003 (2012) [arXiv:1111.4104]; [36] G. Voronov, PoS LATTICE 2012
, 039 (2012) [arXiv:1212.1376] M. Hayakawa, K.-I. Ishikawa, S. Takeda,
and N. Yamada, arXiv:1307.6997; G. Voronov et al. (the LSD collaboration),
in preparation.

\bibitem{thooftlargeN}
G. 't Hooft, Nucl. Phys. B {\bf 72}, 461 (1974), Nucl. Phys. B {\bf 75},
461 (1974).

\bibitem{veneziano}
G. Veneziano, Nucl. Phys B {\bf 117}, 519 (1976).

\end{thebibliography}
\end{document}